\documentclass[a4paper,10pt]{article}

\usepackage{amsthm}
\usepackage{multirow}
\usepackage[english]{babel}
\usepackage{amssymb}
\usepackage{amsfonts}
\usepackage{amsmath}
\usepackage{subfigure}
\usepackage{algorithm}
\usepackage[noend]{algorithmic}
\usepackage{amsmath,amssymb,cite}
\usepackage{epsfig}
\usepackage{pst-node,pstricks}
\usepackage{array}

\usepackage{diagbox}
\usepackage{array}
\usepackage{tabu}
\usepackage{multirow}
\usepackage{pict2e}
\usepackage{slashbox,booktabs,amsmath}
\usepackage{diagbox}
\usepackage{enumitem}
\usepackage{xcolor,colortbl}

\usepackage{graphicx}
\usepackage{psfrag}
\usepackage[nohead, top=2.9cm, bottom=3.2cm, left=2.9cm, right=3.1cm]{geometry}
\usepackage{enumitem}

\newtheorem{theorem}{Theorem}[section]

\newtheorem{corollary}{Corollary}[section]

\newtheorem{definition}{Definition}[section]

\title{Weak and Strong Watermark Numbers Encoded as Reducible Permutation Graph through Self-inverting Permutations\vspace{0.2cm}}

\title{Strong and Weak Watermark Numbers Encoded as Reducible Permutation Graph under Edge Modification Attacks\vspace{0.2cm}}

\title{On the Robustness Against Malicious Attacks of Watermark Numbers Encoded as Reducible Permutation Graphs\vspace{0.2cm}\\
Characterizing Watermark Numbers encoded as Reducible Permutation Graphs against Malicious Attacks\vspace{0.2cm}}

\title{Characterizing Watermark Numbers encoded as Reducible Permutation Graphs against Malicious Attacks\vspace{0.2cm}}

\author{Anna~Mpanti \ \ Stavros~D.~Nikolopoulos \ \ Leonidas~Palios}

\date{}

\begin{document}

\maketitle

\vspace{-0.5cm}

\centerline{\it Department of Computer Science \& Engineering}

\centerline{\it University of Ioannina}

\centerline{\it GR-45110 \ Ioannina, Greece}

\centerline{\tt \{ampanti,stavros,palios\}@cs.uoi.gr}

%----------------------------------------------------------------------

\vskip 0.3in

\begin{abstract}

In the domain of software watermarking, we have proposed several graph theoretic watermarking codec systems for encoding watermark numbers $w$ as reducible permutation flow-graphs $F[\pi^*]$ through the use of self-inverting permutations $\pi^*$. Following up on our proposed methods, we theoretically study the oldest one, which we call W-RPG, in order to investigate and prove its resilience to edge-modification attacks on the flow-graphs $F[\pi^*]$. In particular, we characterize the integer $w\equiv\pi^*$ as strong or weak watermark through the structure of self-inverting permutations $\pi^*$ which encodes it. To this end, for any integer watermark $w \in R_n=[2^{n-1}, 2^n-1]$, where $n$ is the length of the binary representation $b(w)$ of $w$, we compute the minimum number of 01-modifications needed to be applied on $b(w)$ so that the resulting $b(w')$ represents the valid watermark number $w'$; note that a number $w'$ is called valid (or, true-incorrect watermark number) if $w'$ can be produced by the W-RPG codec system and, thus, it incorporates all the structural properties of $\pi^* \equiv w$.

\vspace*{0.1in}
\end{abstract}

\vspace*{0.05in}
%======================================================================================================
\section{Introduction}
\label{sec:Introduction}
%======================================================================================================
%======================================================================================================

Software watermarking is a defense technique used to prevent or discourage software piracy by embedding a signature, that is, an identifier or, equivalently, a watermark representing the owner, in the code \cite{CoNa,Gro}. When an illegal copy is made, the ownership can be claimed by extracting this identifier or watermark. Although watermarking source code is a relatively new field,  a wide range of software watermarking techniques has been proposed, among which the graph-based methods, that encode watermark numbers as graphs whose structure resembles that of real program graphs \cite{Sa,TaNaMoMa}.

Many papers have been appeared in the literature which present and discuss software watermarking techniques and describe current watermarking solutions available in the market, while most of them also discussed possible attacks against watermarking techniques. These attacks are usually made by adversaries who would like to distort the watermark so that once the code is stolen, there is no question of ownership of this intellectual property.

\vspace*{0.05in}
\noindent \textbf{Watermarking.} The software watermarking problem can be described as the problem of embeding a structure $w$ into a program $P$ such that $w$ can be reliably located and extracted from $P$ even after $P$ has been subjected to code transformations such as translation, optimization and obfuscation \cite{CN12,MyCo}. More precisely, given a program $P$, a watermark $w$, and a key $k$, the software watermarking problem can be formally described by the following two functions: $embed(P, w, k)\to P$ and $extract(P_w, k)\to w$.

In the recent years, software watermarking has received considerable attention and many researchers have developed several codec algorithms mostly for watermarks that are encoded as graph-structures \cite{EpGoLaMaMiTo,DaMy,MoCo,Sa,CoTho,CoThoLo}. In this domain, we have proposed several watermarking codec systems for encoding watermark numbers $w$ as reducible permutation flow-graphs $F[\pi^*]$ through the use of self-inverting permutations $\pi^*$. In particular, we have presented an efficient algorithm for encoding a watermark number~$w$ as a self-inverting permutation $\pi^*$ \cite{CN10} and then four algorithms for encoding the self-inverting permutation~$\pi^*$ into a reducible permutation graph~$F[\pi^*]$ whose structure resembles
the structure of real program graphs:
\begin{enumerate}[label=(\roman*)]
\vspace*{-0.12in}
    \item The former of these four algorithms exploits domination relations on
the elements of $\pi^*$ and uses a DAG representation of $\pi^*$ in order to construct the flow-graph $\pi^*$ \cite{CN12}.
    \vspace*{-0.08in}
    \item The second one exploits domination relations on
specific decreasing subsequences of $\pi^*$ \cite{CN12}, while the last two algorithms incorporate important properties which are derived from the bitonic subsequences composing
the self-inverting permutation $\pi^*$ \cite{MN16}.
\end{enumerate}

\vspace*{-0.1in}
\noindent We have also presented efficient decoding algorithms which efficiently extract the number~$w$ from the four reducible permutation graphs~$F[\pi^*]$.
The two main components of our four proposed codec systems, i.e., the self-inverting permutation $\pi^*$ and the reducible permutation graphs~$F[\pi^*]$, incorporate important structural properties which we have claimed that make our systems resilient to attacks.

%\vspace*{0.05in}
%\noindent \textbf{Graph-based Software Watermarking.}

\vspace*{0.05in}
\noindent \textbf{Our Contribution.} In a graph-based software watermarking algorithm, typical attacks can mainly occur in the three ways: edge-modification, edge-insertion or deletion and node-insertion or deletion attacks. Following up on our proposed graph-based software watermarking codec system, in this work we present the valid edge-modification on the edges of $F[\pi^*]$ produced by watermark $w$ following W-RPG codec system. In order to attest its resilience to edge-modification attacks on the main component of this codec system which is the flow-graph $F[\pi^*]$, we demonstrate the minimum valid edge-modification on the edges of $F[\pi^*]$, after which the $F[\pi^*]$ preserves main properties of its structure, produced by watermark $w$ in any range $R=[2^{n-1},2^n-1]$, where $n$ is the length of the binary representation of $w$, but additionally, given an integer watermark number $w$, it can be characterized as strong or weak watermark.

%\vspace*{0.05in}
%\noindent \textbf{Road Map.} The paper is organized as follows: In Section~\ref{sec:Watermarking} we establish the notation and related terminology, we present background results. In Section~\ref{sec:System} we demonstrate a codec system and describe and show that properties of method's components help prevent edge and/or node modifications attacks. In Section~\ref{sec:Strong} we prove the resilience of every watermark $w \in R_n=[2^{n-1},2^n-1]$ and provide the characterizations of the watermark numbers. Finally, in Section~\ref{sec:Concluding} we conclude the paper and discuss possible future extensions.

\vspace*{0.05in}
%======================================================================================================
\section{Graph-based Software Watermarking}
\label{sec:Watermarking}
%======================================================================================================
%======================================================================================================

Software watermarking has received considerable attention and many researchers have developed several codec algorithms mostly for watermarks that are encoded as graph-structures \cite{EpGoLaMaMiTo}. The patent by Davidson and Myhrvold \cite{DaMy} presented the first published software watermarking algorithm. The preliminary concepts of software watermarking also appeared in paper \cite{GhHe} and patents \cite{MoCo,Sa}. Collberg et al. \cite{CoTho,CoThoLo} presented detailed definitions for software watermarking. Authors of papers \cite{ZhYaNiNi,ZhThWa} have given brief surveys of software watermarking research.

Several software watermarking algorithms have been appeared in the literature that encode watermarks as graph structures \cite{CoKoCaTh,CoHuCaToSt,DaMy,VeVaSin}. A wide range of software watermarking techniques has been proposed among which the graph-based methods that encode watermark numbers $w$ as reducible flow-graph structures $F$ capturing such properties which make them resilient to attacks.

Following the graph-based approach, Chroni and Nicolopoulos have been made in recent years an interesting and broad research work on graph-based codec algorithms for encoding watermark numbers $w$ as reducible flow-graphs $F$. Indeed, they extended the class of software watermarking codec algorithms and graph structures by proposing efficient and easily implemented algorithms for encoding numbers as reducible permutation flow-graphs (RPG) through the use of self-inverting permutations (or, for short, SiP). More precisely, they have presented an efficient method for encoding first an integer $w$ as a self-inverting permutation $\pi^*$ and then encoding $\pi^*$ as a reducible permutation flow-graph $F[\pi^*]$ \cite{CN10}; see, also \cite{CNP18}. The watermark graph $F[\pi^*]$ incorporates properties capable to mimic real code, that is, it does not differ from the graph data structures built by real programs. Furthermore, following up on their proposed method, Mpanti et al. \cite{MNR} show the resilience of reducible permutation graph $F[\pi^*]$ under edge-modification with the experimental study of codec algorithms and structure.

Based on this idea and watermarking scheme proposed by these authors, Bento et al. \cite{BBMPS16} introduced a linear-time algorithm which succeeds in retrieving deterministically the n-bit identifiers encoded by such graphs (with $n > 2$) even if $k = 2$ edges are missing. In addition, they proved that $k = 5$ general edge modifications (removals/insertions) can always be detected in polynomial time. Both bounds are tight. Finally, their results reinforce the interest in regarding Chroni and Nikolopoulos's scheme as a possible software watermarking solution for numerous application.

Recently, Mpanti and Nikololopoulos proposed two different reducible permutation flow-graphs, namely,  $F_s[\pi^*]$ and $F_t[\pi^*]$, incorporating important structural properties which are derived from the bitonic subsequences forming the self-inverting permutation $\pi^*$ \cite{MN16}.

\vspace*{0.05in}
%======================================================================================================
\section{The W-RPG Codec System}
\label{sec:System}
%======================================================================================================
%======================================================================================================
In this section we briefly present the codec system, which we shall call W-RPG, proposed by Chroni and Nikolopoulos \cite{CN10, CN12, CNP18}. We firstly discuss the proposed structural components of their model, namely self-inverting permutation (or, for short, SiP) $\pi^*$ and reducible permutation graph (or, for short, RPG) $F[\pi^*]$, and their properties of method�s components, which help prevent edge and/or node modifications attacks.

The codec system W-RPG consists of the algorithms ${\tt Encode\_W.to.SiP}$ and {\tt Decode\_SiP.to.W}, which encode/decode the watermark $w$ into/from a self-inverting permutation $\pi^*$, respectively. Moreover, it includes the encoding algorithm ${\tt Encode\_SiP.to.RPG}$, which encodes the self-inverting permutation $\pi^*$ into a reducible permutation graph $F[\pi^*]$  based on the d-domination relations of the elements of $\pi^*$, and the corresponding decoding algorithm ${\tt Encode\_PRG.to.SiP}$.

\vspace*{0.1in}
%=============================================================================
\subsection{Components}
%=============================================================================
We consider finite graphs with no multiple edges. For a graph~$G$, we denote by $V(G)$ and $E(G)$ the vertex (or, node) set and edge set of $G$, respectively. The \emph{neighborhood}~$N(u)$ of a vertex~$u$ in the graph~$G$ is the set of all the vertices of $G$ which are adjacent to $u$. The \emph{degree} of a vertex~$u$ in the graph~$G$, denoted $deg(u)$, is the number of edges incident on node $u$; thus, $deg(u)=|N(u)|$. For a node $u$ of a directed graph $G$, the number of directed edges coming in $u$ is called the \emph{indegree} of the node $u$, denoted $indeg(u)$, and the number of directed edges leaving $u$ is its outdegree, denoted $outdeg(u)$.

\begin{figure}[t!]
    \hrule\medskip\medskip\smallskip
    \centering
    \includegraphics[scale=0.55]{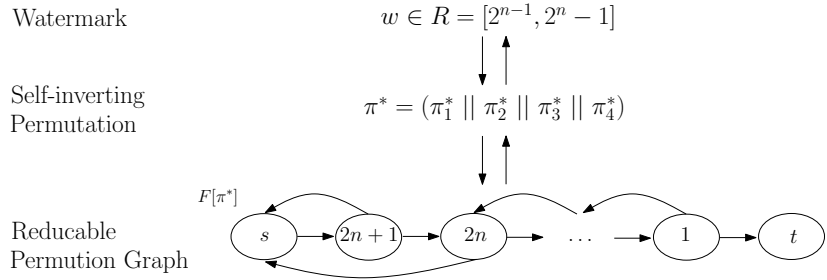}
    \centering
    \smallskip\medskip\hrule\medskip
    \caption{\small{The main data components by algorithms of the codec system for a watermark number $w$.}}
\label{fig:prop}
\end{figure}

The main components, which used by the algorithms of the W-RPG codec system, are illustrated in Figure~\ref{fig:prop} and we describe them in detail.

\vspace*{0.05in}
\noindent \textbf{Self-inverting Permutation.} A permutation $\pi$ over a set $A$ is an arrangement of the elements of the set $A$ into some sequence or order, or if the set $A$ is already ordered, $\pi$ is a rearrangement of the elements of $A$ into a one-to-one correspondence with itself. In this paper, we
consider permutations $\pi$ over the set $N_n = \{1 , 2 , \ldots , n\}$. Let $\pi= ( \pi_1 , \pi_2 , \ldots , \pi_n )$ be such a permutation. By $\pi_i$ we denote the $i$th element of $\pi$, while by $\pi^{-1}_i$ we denote the position in $\pi$ of the element $\pi_i \in N_n$ \cite{Gol80}.

\begin{definition}
Let $\pi=(\pi_1, \pi_2, \ldots, \pi_n)$ be a permutation over the set $N_{n}$, $n>1$. The inverse of the permutation $\pi$ is the permutation $q=(q_1, q_2, \ldots, q_n)$ with $q_{\pi_i} = \pi_{q_i} = i$. A {\it self-inverting permutation} (or, for short, SiP) is a permutation that is its own inverse: $\pi_{\pi_i} = i$.
\end{definition}

\noindent Throughout the paper we shall denote a self-inverting permutation $\pi$ over the set $N_n$ as $\pi^*$.

\vspace*{0.05in}
\noindent \textbf{Reducible Permutation Graph.} A flow-graph is a directed graph $F$ with an initial node $s$ from which all other nodes are reachable. A directed graph $G$ is strongly connected when there is a path $x \rightarrow y$ for all nodes $x$, $y$ in $V(G)$. A node $u \in V(G)$ is an {\it entry} for a subgraph $H$ of the graph $G$ when there is a path $p = (y_1, y_2, \ldots, y_k, u)$ such that $p \cap H = \{u\}$ (see, \cite{HU72,HU74}).

\begin{definition}
A flow-graph is reducible when it does not have a strongly connected subgraph with two (or more) entries.
\end{definition}

\noindent There are some other equivalent definitions of the reducible flow-graphs which use a few more graph-theoretic concepts. A depth first search (DFS) of a flow-graph partitions its edges into tree, forward, back, and cross edges. It is well known that tree, forward, and cross edges form a dag known as a DFS dag. Hecht and Ullman show that a flow-graph $F$ is reducible if and only if $F$ has a unique DFS dag \cite{HU72,HU74}.

\vspace*{0.1in}
%=============================================================================
\subsection{Properties}
\label{subsec:4chain}
%=============================================================================

To be effective, a graph watermark codec system needs to provide several key properties of its structural components. In proposed algorithms by Chroni and Nikolopoulos, the suggested watermarking technique has properties that make it robust to multiple code transformations.

The W-RPG system incorporate several important properties which characterize them as efficient and easily implemented software watermarking systems. Based on the structure of a self-inverting permutation $\pi^*$ produced by Algorithm ${\tt Encode\_W.to.SiP}$, which takes as input an integer $w$, and the type of reducible permutation graphs $F[\pi^*]$, which encoded a elf-inverting permutation $\pi^*$ by Algorithm ${\tt Encode\_SiP.to.RPG-I}$, four important properties of $\pi^*$ or $4-$Chain Property are incorporated into the codec watermark graph $F[\pi^*]$ in order to make it resilient against attacks.

Next we briefly discuss the $4-$Chain Property:

\begin{enumerate}[label=(\roman*)]
\item {\bf SiP Property.} By definition, a permutation is a SiP (self-inverting permutation) if and only if all its cycles are of length 1 or 2.
\item {\bf Bitonic Property}. The self-inverting permutation $\pi^*$ is constructed from the bitonic sequence $\pi^b = X||Y^R$, where $X$ and $Y$ are increasing subsequences and thus the bitonic property of $\pi^b$ is encapsulated in the cycles of $\pi^*$.
\item {\bf Block Property}. Let $B$ be the binary representation of the integer $w$ as input in the algorithm ${\tt Encode\_W.to.SiP}$ and the binary number $B'= 00 \dots 0 ||B||0$. The first part of $B'$ contains the leftmost $n$ bits, each equal to 0, where $n$ is the length of the binary representation of the integer $w$.
\item {\bf Block Property}. The first part of $B'$ contains the leftmost $n$ bits, each equal to 0, where $n$ is the length of the binary representation of the integer $w$.
\item {\bf Range Property}. Let $w$ be the watermark of binary length $n$. The graph $F[\pi^*]$ produced by W-RPG codec system consists of $|V(F[\pi^*])|=n^*+1=2n+3$ nodes, where $n^*=2n+1$.
\end{enumerate}

\noindent Consider a self-inverting permutation $\pi^*$ encoding an integer $w \in R_n=[2^{n-1}, 2^n-1]$, where $n$ is the length of the binary representation of $w$; we distinguish the following two cases:

\vspace*{0.1in}
\noindent \textbf{Zero-and-One case:} $w \in [2^{n-1}, 2^n-2]$. In this case, the structure of $\pi^*$ consists of four subsequences, that is, $\pi^*=\pi_1^* \ || \ \pi_2^* \ || \ \pi_3^* \ || \ \pi_4^*$, having the following forms:
            $$\pi^*=(n+1, n+2, \dots , n+k) \ || \ (p_1, p_2, \ldots, \beta) \ || \ (1,2, \dots , k, \alpha) \ || \ (q_1, q_2, \ldots, \gamma),$$
\noindent where
\begin{enumerate}
    \item[$\circ$] $\pi_{1}^*=(n+1, n+2, \ldots, n+k)$ is an increasing sequence of length $k$ consisting of $k$ consecutive integers starting always with $n+1$, where $k\geq1$,

    \item[$\circ$] $\pi_{2}^*=(p_1, p_2, \ldots, \beta)$ is a bitonic sequence of length $n-k$ with elements of the set $\{n+k+2, n+k+3, \ldots, 2n+1\}$, where $max=2n+1$,

    \item[$\circ$] $\pi_{3}^*=(1, 2, \ldots, k, \alpha)$ is an increasing sequence of length $k+1$ consisting of $k$ consecutive integers starting always with $1$ followed by the integer $\alpha=n+k+1$. Note that, the integer $\alpha$ forms the 1-cycle of $\pi^*$, and

    \item[$\circ$] $\pi_{4}^*=(q_1, q_2, \ldots, \gamma)$ is a sequence of length $n-k$ with elements of the set $\{k+1, k+2, \ldots, k+i, \ldots, n\}$, where $k+i$ is the index of the $i$-th smallest element of $\pi_{2}^*$. Thus, the integer $\gamma$ is the index of the $max=2n+1$ element of $\pi_{2}^*$.
\end{enumerate}

\noindent From the structure of subsequences $\pi_1^*$, $\pi_2^*$, $\pi_3^*$ and $\pi_4^*$ of the SiP $\pi^*$,
it follows that all the element of $\pi_{2}^*$ are greater than that of the sequence $\pi_{1}^*$, the last element~$\beta$ of $\pi_{2}^*$ is greater than any element of the sequence $\pi_3^*||\pi_4^*$, while the last element~$\alpha$ of $\pi_{3}^*$ is greater that any element of $\pi_4^*$.

\vspace*{0.1in}
\noindent \textbf{All-One case:} $w = 2^n-1$. In this case, the sequences $\pi_2^*$ and $\pi_4^*$ have no elements and, thus, $\pi^*$ consists of two subsequences, that is, $\pi^*=\pi_1^* \ || \ \pi_3^*$, having the following forms:
            $$\pi^*=(n+1, n+2, \ldots, 2n) \ || \ (1, 2, \ldots, n, 2n+1),$$
\noindent where
\begin{enumerate}
    \item[$\circ$] $\pi_{1}^*=(n+1, n+2, \ldots, 2n)$ is an increasing sequence of length $n$ consisting of $n$ consecutive integers starting with $n+1$, and

    \item[$\circ$] $\pi_{3}^*=(1, 2, \ldots, n, 2n+1)$ is an increasing sequence of length $n+1$ consisting of $n$ consecutive integers starting always with $1$ followed by the $max=2n+1$ element of $\pi^*$. In this case, the max element $2n+1$ forms the 1-cycle of $\pi^*$.
\end{enumerate}

Furthermore, the reducible permutation flow-graph $F[\pi^*]$ produced by ${\tt Encode\_SiP.to.RPG-I}$ algorithm consists $n^*+2$ nodes, say $u_{n^*+1}$, $u_{n^*}$, $\ldots$, $u_i$, $\ldots$, $u_0$, which include a root node (i.e., every other node in the graph $F[\pi^*]$ is reachable) $s=u_{n^*+1}$ with $outdeg(s)=1$, a footer node (i.e., it is reachable from every other node of the graph) $t=u_0$ with $outdeg(s)=0$ and $n^*$ nodes with two outpointers: one point to node $u_{i+1}$ and the other point to node $u_m$, where $m > i$, where $u_m > u_i > u{_{i-1}}$.

It is worth noting that the reducible permutation flow-graph $F[\pi^*]$ does not differ from the graph data structures built by real programs since its maximum outdegree does not exceed two and it has a unique root node so the program can reach other nodes from the root node, while the self-inverting permutation $\pi^*$ captures important structural properties, due to the Bitonic, Block and Range properties used in the construction of $\pi^*$, which make this codec system resilient to attacks.

\vspace*{0.05in}
%======================================================================================================
\section{Characterization of Watermark Numbers}
\label{sec:Strong}
%======================================================================================================
%======================================================================================================

In this section, we firstly present the way that we can decide, in nearly all cases, whether the reducible permutation graph F produced by W-RPG codec system has suffered an attack on its edges and prove the minimum number of modifications on the edges of $F[\pi^*]$. In the following, we characterize the watermarks $w_i \in R^n$ as weak or strong watermarks, relying on resilience reducible permutation graphs $F[\pi^*]$ to edge modification attacks.

\vspace*{0.1in}
%=============================================================================
\subsection{Edge-Modification Attacks}
%=============================================================================

Let $F[\pi^*]$ be a flow-graph which encodes the integer $w$ and let $F'[*]$ be the graph resulting from $F[\pi^*]$ after an edge modification. Then, we say that $F'[*]$ is either a {\it false-incorrect} or a {\it true-incorrect} graph:

\begin{enumerate}
\item[$\bullet$\,] $F'[*]$ is {\it false-incorrect}
% (or F-incorrect)
if our codec system fails to return
an integer from the graph $F'[*]$, whereas
\item[$\bullet$\,] $F'[*]$ is {\it true-incorrect}
% (or T-incorrect)
if our system extracts from $F'[*]$ and returns
an integer $w' \neq w$.
\end{enumerate}

\noindent
If $F'[*]$ is true-incorrect then there is a SiP $\phi^*$ such that $\phi^* \equiv w'$; in this case $F'[*] = F[\phi^*]$.

Since the SiP properties of the permutation $\pi^*$ which compose the 4-Chain property, i.e., the odd-one property, the bitonic property, the block property, and the range property (see Subsection~\ref{subsec:4chain}) are incorporated in
the structure of the reducible permutation graph~$F[\pi^*]$, it follows that the graph~$F'[\pi^*]$ resulting from $F[\pi^*]$ after any edge modification is false-incorrect if at least one of the SiP properties does not hold.

\begin{definition}
Let $F[\pi^*]$ be a reducible permutation graph produced by watermark $w$ following W-RPG codec system.
We shall call {\it valid edge-modification} a modification on the edges of $F[\pi^*]$, after which the $F[\pi^*]$ preserves the $4-$Chain property; respectively, an {\it invalid edge-modification} leads the decoding process to result on a {\it non-watermark}.
\end{definition}

\vspace*{0.1in}
\noindent
\begin{theorem}   \label{thm:main} Let $w$ be a watermark number encoded as a reducible permutation graph $F[\pi^*]$ through the self-inverting permutation $\pi^*$ and let $w=b_1 b_2 \dots b_n$ be its binary representation and $B=b_2 b_3\dots b_{n-1}$ be the internal block of $w$. For the minimum number of valid edge-modification ${\tt minVM}(w)$ of the graph $F[\pi^*]$ we distinguish the following cases:

\begin{enumerate}
  \item The internal block $B$ of $w$ contains at least two 0s. Then, ${\tt minVM}(w)=3$.

  \item The internal block $B$ of $w$ contains exactly one 0 (i.e., the watermark number has the form $w=1 1^{\ell} 0 1^r b_n$). Then,
    $$
    {\tt minVM}(w)=
    \begin{cases}
    4+min\{\ell ,r-1\},&\text{if \ } b_n=0 \text{ \ and \ } r >0\\
    4, &\text{if \ } b_n=0 \text{ \ and \ } r=0 \\
    4+min\{\ell,r\},& \text{if \ } b_n=1 \text{ \ and \ } r \geq 0
    \end{cases}
    $$
   where $\ell$, $r$ are the numbers of consecutive 1s before and after the 0 in $B$, respectively.

  \item The internal block $B$ of $w$ contains no 0s (i.e., the watermark number has the form $w=1 1 \dots 1 b_n$). Then, ${\tt minVM}(w)=4$.
\end{enumerate}
\end{theorem}

\noindent
{\it Proof.} Let $w$ be a watermark number encoded as a reducible permutation graph $F[\pi^*]$ through the self-inverting permutation $\pi^*$ and let $w=b_1 b_2 \dots b_n$ be its binary representation and $B=b_2 b_3\dots b_{n-1}$ be the internal block of $w$.

\vspace*{0.1in}
\noindent
{\bf Case~1.} The internal block of the watermark number $w$ contains at least two 0s. In this case, by construction (see Algorithm {\tt Encode\_W.to.SiP} \cite{CN10}), the $max=2n+1$ and the $max-1=2n$ elements of $\pi^*$ are not located in the last position of $\pi_{2}^*$, that is, $\beta \neq max-1$ and $\beta \neq max$, or, equivalently, $\pi^{*-1}_{2n} \neq n$ and $\pi^{*-1}_{2n+1} \neq n$.

Let $\gamma=\pi^{*-1}_{2n+1}$ be the index of the $max$ element in $\pi^*$. Since $\pi_{2}^*$ is a bitonic sequence, it follows that the elements $max$ and $max-1$ are in consecutive positions in $\pi_{2}^*$ and thus the index $\pi^{*-1}_{2n}$ of the $max-1=2n$ element is either $\gamma-1$ or $\gamma+1$.

We assume that $\gamma-1=\pi^{*-1}_{2n}$ (the case where $\gamma+1=\pi^{*-1}_{2n}$ is handled in a similar manner). In this case, the watermark number $w$ encodes a SiP $\pi^*=\pi_1^*\ || \ \pi_2^* \ || \ \pi_3^* \ || \ \pi_4^*$ having the following structure:
            \begin{equation}
            \begin{split}
            &\pi^*=(n+1, n+2, \dots , n+k) \ || \ (p_1, p_2, \ldots, p_i, max-1, max, p_j, \ldots, \beta) \ ||\\
            & (1, 2, \dots, k, \alpha) \ || \ (q_1, q_2, \ldots, q_m, \ldots, \gamma-1, \gamma),
            \end{split}
            \end{equation}

\noindent Let $F[\pi^*]$ be the reducible permutation graph which encodes the watermark number $w$ or, equivalently, the self-inverting permutation $\pi^*$. The graph~$F[\pi^*]$ is constructed by computing the function $P(i) = dmax(i)$ for each element $i \in \pi^*$, $1 \leq i \leq 2n+1$; note that, $dmax(i)$ is the maximum element of the set containing all the elements of $\pi^*$ that d-dominate the element~$i$ or, equivalently, $dmax(i)$ is the element with the maximum index in $\pi^*$, which is greater than $i$ and lays on the left of $i$ in $\pi^*$ (see Algorithm {\tt Encode\_SiP.to.RPG} \cite{}). The pair $(dmax(i),i)$, hereafter called {\it d-pair}, forms a back-edge in graph $F[\pi^*]$ from node $i$ to $dmax(i)$; recall that, there is also a forward-edge in $F[\pi^*]$ from node $i$ to $i-1$ ($2 \leq i \leq 2n+1$) and two dummy-nodes $s$ and $t$ such that $(s,2n+1)$ and $(1,t)$ are both forward-edges \cite{CNP18}.

Now we perform valid modifications on the elements of the SiP $\pi^*$. In fact, we apply the process {\tt Swap} on the elements $max-1$ and $max$ of the SiP $\pi^*$ resulting the new SiP $\phi^*$ having the following structure:
            \begin{equation}
            \begin{split}
            &\phi^*=(n+1, n+2, \dots , n+k) \ || \ (p_1, p_2, \ldots, p_i, {max}, {max-1}, p_j, \ldots, \beta) \ ||\\
            & (1, 2, \dots, k, \alpha) \ || \ (q_1, q_2, \ldots, q_m, \ldots, {\gamma}, {\gamma-1}),
            \end{split}
            \end{equation}

\noindent Let $F[\phi^*]$ be the reducible permutation graph with encodes the watermark number $w'\equiv\phi^*$, where $w' \neq w$ since $\phi^* \neq \pi^*$. Both $F[\pi^*]$ and $F[\phi^*]$ have $2n+1$ back-edges, $2n+2$ forward-edges and $2n+3$ nodes. Thus, $F[\phi^*]$ is a true-incorrect reducible permutation graph resulting from $F[\pi^*]$ after performing some edge-modifications on its back-edges.

It is easy to see that the three nodes $max-1$, $p_j$, and $\gamma-1$ are the only true-incorrect nodes of the graphs $F[\pi^*]$ and $F[\phi^*]$. Indeed, the graph $F[\pi^*]$ has 3 back-edges formed by the following d-pairs:
$$(s, max-1), \ \ \ (max, p_j), \ \ \text{and} \ \ \ (q_m, \gamma-1),$$

\noindent where $1 \leq m \leq n-k-2$, while the graph $F[\phi^*]$ has also 3 back-edges formed by the d-pairs:
$$(max, max-1), \ \ \ (max-1, p_j), \ \ \text{and} \ \ \ (\gamma, \gamma-1).$$

From the above, we conclude that the graph $F[\phi^*]$ is a true-incorrect reducible permutation graph, which encodes a watermark number $w' \neq w$, resulting from $F[\pi^*]$ after performing three edge-modifications on its back-edges. Thus, the minimum number of valid edge-modifications in graph $F[\pi^*]$ is 3, that is, ${\tt minVM}(w)=3$.

\vspace*{0.2in}
\noindent
{\bf Case~2.} The internal block $B$ contains exactly one 0 and thus the watermark number has the form:
$$w=1\underbrace{1 \dots 1}_\text{$\ell$}0\underbrace{1 \dots 1}_\text{ $r$} b_n$$
where $\ell \geq 0$ and $r \geq 0$.

\vspace*{0.2in}
\noindent
%======================================================
{\bf Subcase~2.1}: $\mathbf{b_n=0}$ and $\mathbf{r>0}$. In this case, the watermark number $w$ is encoded by a SiP $\pi^*=\pi_1^*||\pi_2^*||\pi_3^*||\pi_4^*$ have the following structure:
            \begin{equation}
            \begin{split}
            &\pi^*=(n+1, n+2, \dots, n+\ell+1) \ || \ (n+\ell+3, \dots, n+\ell+r+2, max, max-1) \ ||\\
            & (1, 2, \dots, \ell+1, \alpha) \ || \ (\ell+2, \dots, \ell+r+1, n, n-1),
            \end{split}
            \end{equation}
\noindent
where $max=2n+1$, $max-1=2n$ and $\alpha=n+\ell+2$. Note that, the watermark number $w$ is of the form
$$w=1\underbrace{1 \dots 1}_\text{$\ell$}0\underbrace{1 \dots 1}_\text{$r $}0$$
\noindent and, thus, $max-1=2n$ is the last element of the sequence $\pi_2^*$;
in fact, $\pi^{-1}_{2n}=n$ and $\pi^{-1}_{2n+1}=n-1$.

Let $\phi^*$ be the SiP resulting from the permutation $\pi^*$ after performing some valid modifications on its elements and let $F[\phi^*]$ be the true-incorrect reducible permutation graph encoding a watermark number $w' \neq w$; the valid modifications belong to the following three categories:

\begin{enumerate}
%---------------------
\item[(i)] {\tt Swap}. Since $\pi_{1}^*$ is an increasing sequence, we cannot apply a {\tt Swap} operation on any pair of elements of $\pi_{1}^*$. Thus, we apply swapping on the elements of $\pi_{2}^*$; note that, the pairs $(max-1,max)$ and $(max-1,max-2)$, where $max-2=n+\ell+r+2$, are the only pairs of elements we can apply such an operation due to the bitonicity of the resulting sequence $\phi_2^*$.

    Applying {\tt Swap()} on the pair $(max,max-1)$, the structure of the resulting SiP $\phi^*$ becomes the following:
            \begin{equation}
            \begin{split}
            &\phi^*=(n+1, n+2, \dots, n+\ell+1) \ || \ (n+\ell+3 ,\dots, n+\ell+r+2, {max-1}, { max}) \ ||\\
            & (1, 2, \dots, \ell+1, \alpha) \ || \ (\ell+2, \dots, \ell+r+1, {n-1}, {n}).
            \end{split}
            \end{equation}

\noindent Since $\phi^*$ is a valid SiP and $\phi^* \neq \pi^*$, the graph $F[\phi^*]$ is a true-incorrect reducible permutation graph encoding the watermark number $w' \neq w$, where
$$w'=1\underbrace{1 \dots 1}_\text{$\ell$}0\underbrace{1 \dots 1}_\text{$r $}1$$
The graph $F[\phi^*]$ has the following true-incorrect nodes:
$$max-1, \ 1, \ 2, \ldots, \ \ell+1, \ \alpha, \ n-1.$$
and thus, in this case, the number of valid edge-modifications in graph $F[\pi^*]$ is $4+\ell$.

Applying now {\tt Swap()} on the pair $(max-2,max-1)$, the graph $F[\pi^*]$ and the resulting true-incorrect reducible permutation graph $F[\phi^*]$ have a number of true-incorrect nodes greater than $4+\ell$. Indeed, in this case the structure of $\phi^*$ becomes the following:
            \begin{equation}
            \begin{split}
            &\phi^*=(n+1, n+2, \dots, n+\ell+1) \ || \ (n+\ell+3, \dots, {max-1}, max, {n+\ell+r+2}) \ ||\\
            & (1, 2, \dots, \ell+1, \alpha) \ || \ (\ell+2, \dots, \ell+r, {n}, {n-2}, n-1)
            \end{split}
            \end{equation}
and, thus, the graph $F[\phi^*]$ contains $5+\ell$ true-incorrect nodes:
$$n+\boldsymbol\ell+r+2, max-1, \ 1, \ 2, \ldots, \ \ell+1, \ \alpha, \ n-2.$$

%--------------------------
\item[(ii)] {\tt Move-in}. We cannot perform any {\tt Move-in} operation on sequence $\pi^*_1$ due to its structure; recall that, $\pi^*_1$ is an increasing sequence consisting of $\ell+1$ consecutive integers starting with $n+1$. Thus, we focus on the sequence $\pi^*_2$ which, in the case we consider, is a bitonic sequence of the form:
    $$\pi^*_2=(n+\ell+3, \dots, n+\ell+i-1, n+\ell+i, n+\ell+i+1, \dots, n+\ell+r+2, max, max-1)$$
    Let $n+\ell+i$ be an element of the increasing subsequence of $\pi^*_2$, where $3 \leq i \leq r+2$. Since $n+\ell+i<max-1$, applying a {\tt Move-in()} operation on the element $n+\ell+i$, the only valid position it can be moved is the last position of $\pi^*_2$. Thus, the resulting sequence $\phi^*_2$ is the following:
     $$\phi^*_2=(n+\ell+3, \dots, n+\ell+i-1, n+\ell+i+1, \dots, n+\ell+r+2, max, max-1, \mathbf{n+\ell+i})$$

Then, $\pi^*_1=\phi^*_1$ and $\pi^*_3=\phi^*_3=(1, 2, \dots, \ell+1, \alpha)$, while the sequences $\pi^*_4$ and $\phi^*_4$ are the following:
    $$\pi^*_4=(\ell+2, \dots, \ell+i-2, \ell+i-1, \ell+i, \dots, \ell+r+1, n, n-1)$$
and
     $$\phi^*_4=(\ell+2, \dots, \ell+i-2, {n}, \ell+i-1, \dots, \ell+r, n-1, n-2)$$

In this case, the resulting true-incorrect graph $F[\phi^*]$ consists of $1+|\phi^*_3|+(r-i+3)$ true-incorrect nodes: the node $n+\ell+i$ of the sequence $\phi^*_2$, all the nodes $1,2, \dots, \ell+1, \alpha$ of the sequence $\phi^*_3$, due to element $n+\ell+i$, and the nodes $\ell+i-1, \dots, \ell+r, n-2$ of the sequence $\phi^*_4$. Thus, the graph $F[\phi^*]$ consists of $6+\ell+r-i$ true-incorrect nodes, where $3 \leq i \leq r+2$ and $r>0$. It follows that, the graph $F[\phi^*]$ can be contain $4+\ell$ true-incorrect nodes.

Note that, $F[\phi^*]$ is a true-incorrect reducible permutation graph resulting from $F[\pi^*]$ after applying a {\tt Move-in()} operation on the element $n+\ell+r+2=max-2$. Moreover, the graph $F[\phi^*]$ encodes the watermark number $w' \neq w$, where
$$w'=1\underbrace{1 \dots 1}_\text{$\ell$}0\underbrace{1 \dots 10}_\text{$r$}0$$

Consider now the case where the {\tt Move-in()} operation is applied on either the element $max$ or $max-1$ of $\pi^*_2$. Both theses cases are reduced to the Case 2.1(i), where the graph $F[\phi^*]$ also contains $4+\ell$ true-incorrect nodes.

%---------------------------
\item[(iii)] {\tt Move-out}. We can perform a {\tt Move-out} operation by moving $i$ elements from $\pi^*_1$ to $\pi^*_3$ and $\pi^*_2$ ($1 \leq i \leq \ell$) or by moving $j$ elements from $\pi^*_2$ to $\pi^*_3$ and $\pi^*_1$ ($1 \leq j \leq r+2$), where
            \begin{equation}
            \begin{split}
            &\pi^*_{1}=(\underbrace{n+1, n+2, \dots, n+\ell+1}_\text{$\ell+1$}) \ \ \pi^*_{2}=(\underbrace{n+\ell+3, \dots, n+\ell+r+2, max, max-1}_\text{$r+2$}) \ \\
            & \pi^*_{3}=(\underbrace{1, 2, \dots, \ell+1, \alpha=n+\ell+2}_\text{$\ell+2$}) \ \ \pi^*_{4}=(\underbrace{\ell+2, \dots, \ell+r+1, n, n-1}_\text{$r+2$}).
            \end{split}
            \end{equation}

(iii.a): We consider first the case where $i$ elements are moved from $\pi^*_1$ to $\pi^*_3$ and $\pi^*_2$, and let $\phi^*$ be the resulting SiP. Based on the structure of both $\pi^*$ and $\phi^*$, such an operation moves the $i$ largest element from $\pi^*_1$ and produces the SiP $\phi^*$ with the following structure:
            \begin{equation}
            \begin{split}
            &\phi^*_{1}=(\underbrace{n+1, \dots, n+\ell+1-i}_\text{$\ell+1-i$}) \ \ \phi^*_{2}=(\underbrace{n+\ell+3-i, \ldots, n+\ell+1, \alpha, n+\ell+3, \dots, max-1}_\text{$r+2+i$}) \ \\
            & \phi^*_{3}=(\underbrace{1, 2, \dots, \ell+1-i, \alpha'=n+\ell+2-i}_\text{$\ell+2-i$}) \ \ \phi^*_{4}=(\underbrace{\ell+2-i, \dots, \ell+1, \ell+2, \dots, n-1}_\text{$r+2+i$}).
            \end{split}
            \end{equation}

The resulting true-incorrect graph $F[\phi^*]$ has the following true-incorrect nodes: the node $\alpha=n+\ell+2$ of the sequence $\phi^*_2$, the node $\alpha'=n+\ell+2-i$ of the sequence $\phi^*_3$, and the $i$ nodes $\ell+2-i, \dots, \ell+1$ plus the $r+1$ nodes $\ell+2, \dots, \ell+r+1, n$ of the sequence $\phi^*_4$, due to element $\alpha'$.
Thus, the graph $F[\phi^*]$ consists of $3+i+r$ true-incorrect nodes, where $1 \leq i \leq \ell$.
It follows that, the graph $F[\phi^*]$ can be contain $4+r$ true-incorrect nodes.

Note that, $F[\phi^*]$ is a true-incorrect reducible permutation graph resulting from $F[\pi^*]$ after applying a {\tt Move-out()} operation on the element $n+\ell+1$ of $\pi^*_{1}$. The true-incorrect graph $F[\phi^*]$ encodes the watermark number $w' \neq w$ of the form:
$$w'=1\underbrace{1 \dots 1}_\text{$\ell-i$}0\underbrace{1 \dots 1}_\text{$r+i$}0$$

(iii.b): Consider now the case where $j$ elements are moved from $\pi^*_2$ to $\pi^*_3$ and $\pi^*_1$, where $1 \leq j \leq r$; the cases where $j=r+1$ and $j=r+2$ will be considered separately. Let $\phi^*$ be the resulting SiP after a moving operation. Again, based on the structure of both permutations $\pi^*$ and $\phi^*$, the moved elements are the $j$ smallest elements $n+\ell+3, n+\ell+4, \dots, n+\ell+2+j$ of $\pi^*_2$ and the resulting SiP $\phi^*$ has the following structure:
            \begin{equation}
            \begin{split}
            &\phi^*_{1}=(\underbrace{n+1, \dots, n+\ell+1, \alpha, n+\ell+3, \ldots, n+\ell+1+j}_\text{$\ell+1+j$}) \ \ \phi^*_{2}=(\underbrace{n+\ell+3+j, \dots,  max-1}_\text{$r+2-j$}) \ \\
            & \phi^*_{3}=(\underbrace{1, 2, \dots, \ell+1+j, \alpha'=n+\ell+2+j}_\text{$\ell+2+j$}) \ \ \phi^*_{4}=(\underbrace{\ell+2+j, \dots, n, n-1}_\text{$r+2-j$}).
            \end{split}
            \end{equation}

The graph $F[\phi^*]$ contains of the following true-incorrect nodes: the node $\alpha=n+\ell+2$ of the sequence $\phi^*_1$, the node $\alpha'=n+\ell+2+j$ of the sequence $\phi^*_3$, and the $j$ nodes $\ell+2, \ell+3, \dots, \ell+1+j$ plus the $r+1-j$ nodes $\ell+2+j, \dots, n$ of the sequence $\phi^*_4$, due to element $\alpha'$.
Thus, the graph $F[\phi^*]$ consists of $3+r$ true-incorrect nodes and encodes the watermark number $w' \neq w$ of the form:
$$w'=1\underbrace{1 \dots 1}_\text{$\ell+j$}0\underbrace{1 \dots 1}_\text{$r-j$}0$$
where, $j=1, 2, \ldots, r$.

The case where the $j=r+1$ smallest elements of $\pi^*_2$ are moved to $\pi^*_3$ and $\pi^*_1$ is handled as follows: the resulting SiP $\phi^*$ has the structure
            \begin{equation}
            \begin{split}
            &\phi^*_{1}=(n+1, n+2, \ldots, n+\ell+1, \alpha, \ldots, max-2=2n-1) \ \ \phi^*_{2}=(max) \ \\
            & \phi^*_{3}=(1, 2, \dots, \ell+1, \ell+2, \ldots, n-1, \alpha'=max-1) \ \ \phi^*_{4}=(n)
            \end{split}
            \end{equation}
and, thus, the corresponding true-incorrect graph $F[\phi^*]$ contains one true-incorrect node in $\phi^*_{1}$ (i.e., the node $\alpha$), $n-1$ true-incorrect nodes in $\phi^*_{3}$ (i.e., the nodes $1, 2, \ldots, n-1$) and one true-incorrect node in $\phi^*_{4}$ (i.e., the node $n$). In this case, the graph $F[\phi^*]$ consists of $n+1$ true-incorrect nodes, where $n\geq4$, and encodes the watermark number $w' \neq w$ of the form:
$$w'=1\underbrace{11 \dots 1}_\text{$n-2$}0$$

In the case where $j=r+2$ elements of $\pi^*_2$ are moved to $\pi^*_3$ and $\pi^*_1$ (i.e., all the elements of $\pi^*_2$ are moved), the sequences $\phi^*_2$ and $\phi^*_4$ of the resulting SiP $\phi^*$ are empty. Thus, $\phi^*$ actually consists of two increasing sequences $\phi^*=\phi^*_1 || \phi^*_3$ and has the following structure:
            \begin{equation}
            \begin{split}
            &\phi^*_{1}=(n+1, n+2, \ldots, n+\ell+1, \alpha, \ldots, max-1=2n) \ \ \phi^*_{2}=() \ \\
            & \phi^*_{3}=(1, 2, \dots, \ell+1, \ell+2, \ldots, n, \alpha'=max=2n+1) \ \ \phi^*_{4}=().
            \end{split}
            \end{equation}
The resulting graph $F[\phi^*]$ contains two true-incorrect nodes in $\phi^*_1$, i.e., the nodes $\alpha=n+\ell+2$ and $max-1=2n$, and $r+2=|\phi^*_{2}|$ true-incorrect nodes in $\phi^*_2$, i.e., the nodes $\ell+2, \ell+3, \dots, n$, due to element $max-1$.
Thus, the graph $F[\phi^*]$ consists of $4+r$ true-incorrect nodes and encodes the watermark number $w' \neq w$ of the form:
$$w'=1\underbrace{11 \dots 1}_\text{$n-2$}1$$

\end{enumerate}

\noindent Summing up the case where $b_n=0$ and $r>0$, we conclude that a true-incorrect reducible permutation graph $F[\phi^*]$ encoding a watermark number $w' \neq w$ can be result from $F[\pi^*]$ after performing either $4+\ell$ or $3+r$ edge-modifications on its back-edges. Thus, ${\tt minVM}(w)=4+min\{\ell ,r-1\}$.

\vspace*{0.2in}
\noindent
{\bf Subcase~2.2}: $\mathbf{b_n=0}$ and $\mathbf{r=0}$. In this case, the watermark number $w$ is encoded by a SiP $\pi^*=\pi_1^*||\pi_2^*||\pi_3^*||\pi_4^*$ having the following structure:
            \begin{equation}
            \begin{split}
            &\pi^*=(n+1, n+2, \dots, n+\ell+1) \ || \ (max, max-1) \ ||\\
            & (1, 2, \dots, \ell+1, \alpha=n+\ell+2) \ || \ (n, n-1),
            \end{split}
            \end{equation}
\noindent
where $max=2n+1$, $max-1=2n$ and $\ell=n-3$. The watermark number $w$ is of the following form:
$$w=1\underbrace{111 \dots 1}_\text{$\ell=n-3$}00$$
\noindent Let $\phi^*$ be the SiP resulting from $\pi^*$ after performing some valid modifications on its elements. As in the previous case, we consider valid modifications by performing the operations Swap(), Move-in() and Move-out() on the elements of $\pi^*$.

\begin{enumerate}
%---------------------
\item[(i)] {\tt Swap}. Recall that we can apply swapping only on the elements of the bitonic sequence $\pi_{2}^*$. Thus, applying {\tt Swap()} on the pair $(max,max-1)$, the structure of the resulting SiP $\phi^*$ becomes the following:
            \begin{equation}
            \begin{split}
            &\phi^*=(n+1, n+2, \dots, n+\ell+1) \ || \ ({max-1}, {max}) \ ||\\
            & (1, 2, \dots, \ell+1, \alpha=n+\ell+2) \ || \ ({n-1}, {n}).
            \end{split}
            \end{equation}

The true-incorrect graph $F[\phi^*]$ encoding the SiP $\phi^*$ contains one true-incorrect node in $\phi^*_2$, i.e., the node $max-1$, one true-incorrect node in $\phi^*_4$, i.e., the node $n-1$ and $\ell+2$ true-incorrect nodes in $\phi^*_3$, i.e., the nodes $1, 2, \dots, \ell+1, \alpha$. Thus, in total, in this case the graph $F[\phi^*]$ contains $4+\ell$ true-incorrect nodes.

%--------------------------
\item[(ii)] {\tt Move-in}. It is easy to see that the SiP $\phi^*$ which results from $\pi^*$ by applying a moving-in operation on sequence $\pi^*_2$ has the same structure with that which results in the previous Case 2.2(i). Thus, the graph $F[\phi^*]$ contains $3+\ell=n$ true-incorrect nodes, where $n\geq4$.

%---------------------------
\item[(iii)] {\tt Move-out}. In this case, we can perform a {\tt Move-out} operation either by moving $i$ elements from $\pi^*_1$ to $\pi^*_3$ and $\pi^*_2$ ($1 \leq i \leq \ell$) or, since $r=0$, by moving $j$ elements from $\pi^*_2$ to $\pi^*_3$ and $\pi^*_1$ ($1 \leq j \leq 2$), where
            \begin{equation}
            \begin{split}
            &\pi^*_{1}=(\underbrace{n+1, n+2, \dots, n+\ell+1}_\text{$\ell+1$}) \ \ \pi^*_{2}=(\underbrace{max, max-1}_\text{$2$}) \ \\
            & \pi^*_{3}=(\underbrace{1, 2, \dots, \ell+1, \alpha=n+\ell+2}_\text{$\ell+2$}) \ \ \pi^*_{4}=(\underbrace{n, n-1}_\text{$2$}).
            \end{split}
            \end{equation}

(iii.a): We consider first the case where $i$ elements are moved from $\pi^*_1$ to $\pi^*_3$ and $\pi^*_2$, and let $\phi^*$ be the resulting SiP. This case results from the Case 2.1(iii) by setting $r=0$. It follows that, there exists a graph $F[\phi^*]$ having exactly $4$ true-incorrect nodes, that is, $F[\phi^*]$ can be constructed from $F[\pi^*]$ by modifying exactly $4$ edges.

The watermark number $w' \neq w$ encoded by the true-incorrect graph $F[\phi^*]$ is of the form:
$$w'=1\underbrace{1 \dots 1}_\text{$\ell-1$}010$$

(iii.b): The cases where the $max-1$ element or both the $max-1$ and $max$ elements are moved from $\pi^*_2$ result to a true-incorrect graph $F[\phi^*]$ consisting of more than 4 true-incorrect nodes.

\end{enumerate}

\noindent Summarizing the results of the case where $b_n=0$ and $r=0$, we conclude that a true-incorrect reducible permutation graph $F[\phi^*]$ encoding a watermark number $w' \neq w$ can be result from $F[\pi^*]$ after performing at least $4$ edge-modifications on its back-edges. Thus, ${\tt minVM}(w)=4$.

%=======================================================================================
%=======================================================================================

\vspace*{0.1in}
\noindent
{\bf Subcase~2.3}: $\mathbf{b_n=1}$ and $\mathbf{r\geq0}$.
In this case, the watermark number $w$ is encoded by a SiP $\pi^*=\pi_1^* \ || \ \pi_2^* \ || \ \pi_3^* \ || \ \pi_4^*$ have the following structure:
            \begin{equation}
            \begin{split}
            &\pi^*=(n+1, n+2, \dots, n+\ell+1) \ || \ (n+\ell+3, \dots, n+\ell+r+2, max-1, max) \ ||\\
            & (1, 2, \dots, \ell+1, \alpha) \ || \ (\ell+2, \dots, \ell+r+1, n-1, n),
            \end{split}
            \end{equation}
\noindent
where $\alpha=n+\ell+2$. The $max=2n+1$ element is located the last position of the sequence $\pi^*_2$ and, thus, $\pi^*_2$ is an increasing sequence of length $r+2$. In the case under consideration, the watermark number $w$ encoded by $\pi^*$ is of the form:
$$w=1\underbrace{1 \dots 1}_\text{$\ell$}0\underbrace{1 \dots 1}_\text{$r $}1$$

\noindent
Let $\phi^*$ be the SiP resulting from $\pi^*$ after performing some valid modifications on its elements and let $F[\phi^*]$ be the true-incorrect reducible permutation graph encoding a watermark number $w' \neq w$. As in the previous cases, the SiP $\phi^*$ is produced by applying the three operations Swap(), Move-in() and Move-out() on the elements of $\pi^*$.

\begin{enumerate}
%---------------------
\item[(i)] {\tt Swap}. We can apply swapping only on the elements of $\pi_{2}^*$; in fact, the pair $(max-1,max)$ is the only pair of elements of $\pi_{2}^*$ we can apply a swap operation due to the bitonicity of the resulting sequence $\phi_2^*$.

    After applying the operation {\tt Swap()} on the pair $(max-1,max)$, the structure of the resulting SiP $\phi^*$ becomes the following:
            \begin{equation}
            \begin{split}
            &\phi^*=(n+1, n+2, \dots, n+\ell+1) \ || \ (n+\ell+3 ,\dots, n+\ell+r+2, {max}, {max-1}) \ ||\\
            & (1, 2, \dots, \ell+1, \alpha) \ || \ (\ell+2, \dots, \ell+r+1, {n}, {n-1}).
            \end{split}
            \end{equation}

\noindent
The graph $F[\phi^*]$ is a true-incorrect reducible permutation graph encoding the watermark number $w' \neq w$, where
$$w'=1\underbrace{1 \dots 1}_\text{$\ell$}0\underbrace{1 \dots 1}_\text{$r $}0$$
and has the following true-incorrect nodes:
$$max-1, \ 1, \ 2, \ldots, \ \ell+1, \ \alpha, \ n-1.$$
Thus, in this case, the number of valid edge-modifications in graph $F[\pi^*]$ is $4+\ell$.

%--------------------------
\item[(ii)] {\tt Move-in}. We can perform any moving-in operation only on the sequence $\pi^*_2$ which, in the case we consider, is a increasing sequence of the form:
    $$\pi^*_2=(n+\ell+3, n+\ell+4, \dots, n+\ell+r+2, max-1, max)$$
The sequence $\pi^*_4$, the elements of which are the indices of the elements of $\pi^*_2$ in $\pi^*$, has the following form:
    $$\pi^*_4=(\ell+2, \ell+3, \dots, \ell+r+1, n-1, n)$$

\noindent
Let $p_1, p_2, \ldots, p_i$ be $i$ elements of $\pi^*_2$ such that $p_1 < p_2 < \cdots < p_i$, where $p_1=n+\ell+m, 3 \le m \le r+3$. We perform moving-in operations on the elements $p_1, p_2, \ldots, p_i$ and let $\phi^*_2$ be the resulting bitonic sequence of the resulting SiP $\phi^*_2$. Then, $\phi^*_2$ has the following form:
    $$\phi^*_2=({\tt P}, max, p_i, p_{i-1}, \ldots, p_1)$$
where ${\tt P}$ is an increasing sequence of length $r+1-i$ consisting of the remaining elements of $\pi^*_2$ lying on the left of $max$ after applying the moving-in operations. Let $q_1, q_2, \ldots, q_j$ be the elements $\pi^{*-1}_{p_1}, \pi^{*-1}_{p_1}+1, \ldots , \pi^{*-1}_{max-1}$. Then, the sequence $\phi^*_4$ has the form:
    $$\phi^*_4=({\tt Q}, n, q_1, q_{2}, \ldots, q_j)$$
where {\tt Q} is an increasing sequence of length $m-3$ consisting of the indices of the elements of $(n+\ell+3, n+\ell+4, \ldots, n+\ell+m-1)$. Moreover, $\pi^*_1=\phi^*_1=(n+1, n+2, \dots, n+\ell+1)$ and $\pi^*_3=\phi^*_3=(1, 2, \dots, \ell+1, \alpha)$.

Thus, the corresponding true-incorrect graph $F[\phi^*]$ has the following true-incorrect nodes: the nodes $p_i, p_{i-1}, \ldots, p_1$ of the sequence $\phi^*_2$, the nodes $1, 2, \dots, \ell+1, \alpha$ of the sequence $\phi^*_3$, and the nodes $q_1, q_{2}, \ldots, q_j$ of the sequence $\phi^*_4$. In total, the graph $F[\phi^*]$ contains $i+\ell+2+( \pi^{*-1}_{max}-\pi^{*-1}_{p_1} ) = i+\ell+2+(n-m)$ true-incorrect nodes. By getting $i=1$ and $(n-m)=1$, we conclude that there exist a graph $F[\phi^*]$ containing $4+\ell$ true-incorrect nodes or, equivalently, we can obtain a true-incorrect graph $F[\phi^*]$ after performing $4+\ell$ edge-modifications in graph $F[\pi^*]$.

\item[(iii)] {\tt Move-out}. We perform a {\tt Move-out} operation either by moving $i$ elements from $\pi^*_1$ to $\pi^*_3$ and $\pi^*_2$ ($1 \leq i \leq \ell$) or by moving $j$ elements from $\pi^*_2$ to $\pi^*_3$ and $\pi^*_1$ ($1 \leq j \leq r+2$), where
            \begin{equation}
            \begin{split}
            &\pi^*_{1}=(\underbrace{n+1, n+2, \dots, n+\ell+1}_\text{$\ell+1$}) \ \ \pi^*_{2}=(\underbrace{n+\ell+3, \dots, n+\ell+r+2, max-1, max}_\text{$r+2$}) \ \\
            & \pi^*_{3}=(\underbrace{1, 2, \dots, \ell+1, \alpha=n+\ell+2}_\text{$\ell+2$}) \ \ \pi^*_{4}=(\underbrace{\ell+2, \dots, \ell+r+1, n-1, n}_\text{$r+2$}).
            \end{split}
            \end{equation}

(iii.a): We consider first the case where $i$ elements are moved from $\pi^*_1$ to $\pi^*_3$ and $\pi^*_2$, and let $\phi^*$ be the resulting SiP. Such an operation moves the $i$ largest element from $\pi^*_1$ and produces the SiP $\phi^*$ with the following structure:
            \begin{equation}
            \begin{split}
            &\phi^*_{1}=(\underbrace{n+1, \dots, n+\ell+1-i}_\text{$\ell+1-i$}) \ \ \phi^*_{2}=(\underbrace{n+\ell+3-i, \ldots, n+\ell+1, \alpha, n+\ell+3, \dots, max}_\text{$r+2+i$}) \ \\
            & \phi^*_{3}=(\underbrace{1, 2, \dots, \ell+1-i, \alpha'=n+\ell+2-i}_\text{$\ell+2-i$}) \ \ \phi^*_{4}=(\underbrace{\ell+2-i, \dots, \ell+1, \ell+2, \dots, n}_\text{$r+2+i$}).
            \end{split}
            \end{equation}

The resulting true-incorrect graph $F[\phi^*]$ has of the following true-incorrect nodes: the node $\alpha=n+\ell+2$ of the sequence $\phi^*_2$, the node $\alpha'=n+\ell+2-i$ of the sequence $\phi^*_3$, and the $i$ nodes $\ell+2-i, \dots, \ell+1$ plus the $r+2$ nodes $\ell+2, \dots, \ell+r+1, n-1, n$ of the sequence $\phi^*_4$.
Thus, the graph $F[\phi^*]$ consists of $4+i+r$ true-incorrect nodes, where $1 \leq i \leq \ell$.
It follows that, there exist a graph $F[\phi^*]$ that can be contain $5+r$ true-incorrect nodes.

(iii.b): We follow a similar approach as in the Case 2.1(iii.b). Let $j$ elements are moved from $\pi^*_2$ to $\pi^*_3$ and $\pi^*_1$, where $1 \leq j \leq r$; the cases where $j=r+1$ and $j=r+2$ will be considered separately. The moved elements are the $j$ smallest elements $n+\ell+3, n+\ell+4, \dots, n+\ell+2+j$ of $\pi^*_2$ and the resulting SiP $\phi^*$ has the following structure:
            \begin{equation}
            \begin{split}
            &\phi^*_{1}=(\underbrace{n+1, \dots, n+\ell+1, \alpha, n+\ell+3, \ldots, n+\ell+1+j}_\text{$\ell+1+j$}) \ \ \phi^*_{2}=(\underbrace{n+\ell+3+j, \dots,  max}_\text{$r+2-j$}) \ \\
            & \phi^*_{3}=(\underbrace{1, 2, \dots, \ell+1+j, \alpha'=n+\ell+2+j}_\text{$\ell+2+j$}) \ \ \phi^*_{4}=(\underbrace{\ell+2+j, \dots, n-1, n}_\text{$r+2-j$}).
            \end{split}
            \end{equation}

The graph $F[\phi^*]$ contains the following true-incorrect nodes: the node $\alpha=n+\ell+2$ of the sequence $\phi^*_1$, the node $\alpha'=n+\ell+2+j$ of the sequence $\phi^*_3$, and the $j$ nodes $\ell+2, \ell+3, \dots, \ell+1+j$ plus the $r+2-j$ nodes $\ell+2+j, \dots, n-1, n$ of the sequence $\phi^*_4$, due to element $\alpha'$.
Thus, the graph $F[\phi^*]$ consists of $4+r$ true-incorrect nodes and encodes the watermark number $w' \neq w$ of the form:
$$w'=1\underbrace{1 \dots 1}_\text{$\ell+j$}0\underbrace{1 \dots 1}_\text{$r-j$}1$$
where, $j=1, 2, \ldots, r$.

The cases where the $j=r+1$ and $j=r+2$ are exactly the same as the corresponding cases in Case 2.1(iii.b).

\end{enumerate}

\noindent Concluding the case where $b_n=1$ and $r \geq 0$, it holds that a true-incorrect reducible permutation graph $F[\phi^*]$ encoding a watermark number $w' \neq w$ can be result from $F[\pi^*]$ after performing either $4+\ell$ or $4+r$ edge-modifications on its back-edges. Thus, ${\tt minVM}(w)=4+min\{\ell ,r\}$.

\vspace*{0.2in}
\noindent
{\bf Case~3.} In this case, the internal block $B$ of the watermark number $w$ contains no 0s and, thus, the binary representation of the number $w$ has one of the following two forms:
$$1\underbrace{1 \dots 11}_\text{$n-2$}0 \ \ \ \ \ \text{or} \ \ \ \ \ 1\underbrace{1 \dots 11}_\text{$n-2$}1$$

\begin{enumerate}
\item[(3.1)] In the former case where $b_n=0$, the watermark number $w$ is encoded by a SiP $\pi^*=\pi_1^* \ || \ \pi_2^* \ || \ \pi_3^* \ || \ \pi_4^*$ having the following structure:
\begin{equation}
\begin{split}
&\pi^*=(n+1, n+2, \dots, 2n-i-1, 2n-i, \ldots, 2n-1) \ || \ (max) \ ||\\
& (1, 2, \dots, n-i-1, n-i, \ldots, n-1, \alpha=2n) \ || \ (n),
\end{split}
\end{equation}
\noindent
where $max=2n+1$.

From the structures of $\pi^*_1$ and $\pi^*_2$, it follows that the only operation we can apply is the {\tt Move-out} on the elements of both theses sequences. Let first consider the case where
$i$ largest elements are moved from $\pi^*_1$, i.e., the elements $2n-i, 2n-i+1, \ldots, 2n-1$. Then, the structure of the resulting SiP $\phi^*$ becomes the following:
\begin{equation}
\begin{split}
&\phi^*=(n+1, n+2, \dots, 2n-i-1) \ || \ (2n-i+1, \ldots, 2n-1, \alpha, max) \ ||\\
& (1, 2, \dots, n-i-1, \alpha'=2n-i) \ || \ (n-i, \ldots, n-1, n),
\end{split}
\end{equation}

\noindent The resulting true-incorrect graph $F[\phi^*]$ has of the following true-incorrect nodes: the node $\alpha=2n$ in sequence $\phi^*_2$, the node $\alpha'=2n-i$ in sequence $\phi^*_3$, and the $i+1$ nodes $n-i, \ldots, n-1, n$ in sequence $\phi^*_4$. Thus, in total the graph $F[\phi^*]$ has $2+i+1$ true-incorrect nodes. Since $1 \leq i \leq n-1$, it follows that the graph $F[\phi^*]$ can be obtain from $F[\pi^*]$ by modifying at least $4$ edges.

The watermark number $w' \neq w$ encoded by a true-incorrect graph $F[\phi^*]$ resulting from $F[\pi^*]$ after exactly $4$ edges modifications, has the following binary form:
$$w'=1\underbrace{1 \dots 10}_\text{$n-2$}1$$

Let us now consider the case where the element $max$ is moved from $\pi^*_2$. Then, both sequences $\phi^*_2$ and $\phi^*_4$ become empty and the sequences $\phi^*_1$ and $\phi^*_3$ have the following structure:
\begin{equation}
\begin{split}
&\phi^*=(n+1, n+2, \dots, 2n-i-1, 2n-i, \ldots, 2n-1, \alpha=2n) \ || \ () \ ||\\
& (1, 2, \dots, n-i-1, n-i, \ldots, n-1, \alpha'=max) \ || \ (),
\end{split}
\end{equation}

\noindent The incorrect nodes of the graph $F[\phi^*]$ are the following: $\alpha'=max$, $\alpha=2n$ and the nodes $1, 2, \dots, n-i-1, n-i, \ldots, n-1$ of $\phi^*_3$, due to the last element $\alpha=2n$ of $\phi^*_1$. Thus, the graph $F[\phi^*]$ can be obtain from $F[\pi^*]$ by modifying $n+1$ edges, where $n \geq 4$.

\item[(3.2)] In the latter case where $b_n=1$, the watermark number $w$ is encoded by a SiP $\pi^*=\pi_1^* \ || \ \pi_2^* \ || \ \pi_3^* \ || \ \pi_4^*$ having the following structure:
\begin{equation}
\begin{split}
&\pi^*=(n+1, n+2, \dots, 2n-i, 2n-i+1, \ldots, 2n) \ || \ () \ ||\\
& (1, 2, \dots, n-i-1, n-i, \ldots, n-1, n, \alpha=max) \ || \ (),
\end{split}
\end{equation}
\noindent
where $max=2n+1$.

Since $\pi^*_1$ is an increasing sequence, the only operation we can apply on $\pi^*_1$ is the {\tt Move-out} operation.
Let the $i$ largest elements of $\pi^*_1$, i.e., $2n-i+1, \ldots, 2n-1, 2n$, are moved to $\pi^*_2$ and $\pi^*_3$. Then, by choosing the last element $2n$ of $\pi^*_1$ to be the last element of $\pi^*_2$, the structure of the resulting SiP $\phi^*$ becomes the following:
\begin{equation}
\begin{split}
&\phi^*=(n+1, n+2, \dots, 2n-i) \ || \ (2n-i+2, \ldots, \alpha=max, 2n) \ ||\\
& (1, 2, \dots, n-i, \alpha'=2n-i+1) \ || \ (n-i+1, \ldots, n, n-1),
\end{split}
\end{equation}

\noindent \noindent In this case, the true-incorrect nodes of the graph $F[\phi^*]$ are the following: the nodes $2n$ and $\alpha'=2n-i+1$ of $\phi^*_2$ and $\phi^*_3$, respectively, the nodes $n-i+1, \ldots, n$, due to the last element $\alpha'=2n-i+1$ of $\phi^*_3$, and the node $n-1$, due to the node $n$ of $\phi^*_4$. Thus, the graph $F[\phi^*]$ can be obtain from $F[\pi^*]$ by modifying $2+i$ edges. Since we require the element $2n$ to be in the last position of $\pi^*_2$, we have that $i \geq 2$. Thus, the graph $F[\phi^*]$ can be obtain from $F[\pi^*]$ by modifying $4$ edges and the watermark number $w' \neq w$ encoded by $F[\phi^*]$ has the following binary form:
$$w'=1\underbrace{1 \dots 10}_\text{$n-2$}0$$

It is easy to see that in the case where the element $2n$ is not located in the last position of $\pi^*_2$, the number of the true-incorrect nodes of the graph $F[\phi^*]$ is greater than 4.

\end{enumerate}

\noindent Summarizing the results of the Case~3, we conclude that a true-incorrect reducible permutation graph $F[\phi^*]$ can be result from $F[\pi^*]$, which encodes the number $w$, after performing at least $4$ edge-modifications on its back-edges. Thus, ${\tt minVM}(w)=4$. \ $\blacksquare$

%====================================================================================================

%\vspace*{0.1in}
%\noindent
%\begin{theorem}
%Let $\pi^*$ be a self-inverting permutation. Given the first increasing subsequence of  $\pi^*$, we can construct the SiP $\pi^*=\pi_1^* \ || \ \pi_2^* \ || \ \pi_3^* \ || \ \pi_4^*$.
%\end{theorem}
%
%\noindent
%{\it Proof.} Let $\pi_{inc}^*=\pi_1^* \ || \ \pi_{21}^*$, where $\pi_{1}^*=(n+1, n+2, \ldots, n+k)$ is an increasing sequence of length $k$ consisting of $k$ consecutive integers starting always with $n+1$, where $k\geq1$, $\pi_{2}^*=\pi_{21}^* \ || \ \pi_{22}^*$ and $\pi_{21}^*=(p_1, p_2, \ldots,max)$ is also an increasing sequence where $max=2n+1$; note that, $\pi^*$ encodes a watermark number $w$ of binary length $n$ and the length of SiP $\pi^*$ is $n^*=2n+1$.
%Then, we can construct the decreasing subsequence $\pi^*_{22}$, $\pi^*_{22} \in [n+1,2n]$.
%
%Also, we know the indices $\pi^{-1}_i$, where $1 \leq i \leq max$ and $max=2n+1$. We can compute which element
%is the 1-cycle $\alpha=n+k+1$ if we know the first increasing subsequence $\pi_1^*$ of $\pi^*$, which is the elements
%between $n + 1$ to $2n + 1$, where $n$ is the binary length of the watermark number $w$.
%
%Thus, the self-inverting permutation $\pi^*$ can be reconstructed by either the first increasing subsequence $\pi_{inc}^*$% or $\pi^{*\beta}$ (bitonic sequence) or $\pi^{*\gamma}$ (index sequence)
%. \ $\blacksquare$

\vspace*{0.1in}
%=============================================================================
\subsection{Strong and Weak Watermarks}
%=============================================================================

We next define the integer $w_i \in R_n =[2^{n-1}, 2^n-1]$ as weak or strong watermark, encoded as a reducible permutation graph $F[\pi^*]$ through the self-inverting permutation $\pi^*$.

%\vspace*{0.15in}
%\noindent {\bf Definition~}
\begin{definition}
A watermark $w\in R_n$ is called {\it strong} if it has $\max\{minVM(w)\}$ in the range $R_n$
and the minimum number of true-watermarks $w'$.
\end{definition}

%==============================================================================================================

\begin{corollary}
Let $w$ be an integer in $R_n=[2^{n-1}, 2^n-1]$.
\begin{enumerate}[label=(\it \roman*)]
  \item If $n=2\kappa + 1,\kappa \in \mathbb{Z}$ then the strong watermark $w$ is of the form $w=1 1^{\ell} 0 1^{\ell} 1$.
  \item If $n=2\kappa,\kappa \in \mathbb{Z}$ then the strong watermark $w$ is of the form $w=1 1^{\ell} 0 1^{\ell+1} 1$.

\end{enumerate}
\end{corollary}

\vspace*{0.1in}

\noindent
{\it Proof.} Let $w$ be an integer in $R_n=[2^{n-1}, 2^n-1]$ where $n$ is the length of the binary representation of $w$. The maximum value of minimum valid edge-modifications $max\{{\tt minVM}(w)\}$ in the range $R_n$ is $4+min\{\ell ,r-1\}$, if $b_n=0$ and $r >0$ or $4+min\{\ell,r\}$, if $b_n=1$ and $r \geq 0$.

\begin{enumerate}[label=(\it \roman*)]
  \item If $n=2\kappa + 1,\kappa \in \mathbb{Z}$ then it is easy to see by the Theorem~\ref{thm:main} that the binary representation of watermark $w$ is unique in the range $R_n$ and is of the form $w=1 1^{\ell} 0 1^r b_n$, where $\ell=r$ and $b_n=1$ and $max\{{\tt minVM}(w)\}=4+\ell$.
  \item If $n=2\kappa,\kappa \in \mathbb{Z}$ then the binary representation of watermark $w$ is of the form $w=1 1^{\ell} 0 1^r b_n$, where $r-\ell=1$ and $b_n=0$, with $max\{{\tt minVM}(w)\}=4+\ell$ or $|r-\ell|=1$ and $b_n=1$, with $max\{{\tt minVM}(w)\}=4+\ell$ or $max\{{\tt minVM}(w)\}=4+r$. It means that there are three integers $w$ in the range $R_n$ with $max\{{\tt minVM}(w)\}$. Hence, the strong watermark $w$ is defined as the watermark which have the minimum number of true-watermarks $w'$ with these minimum valid edge-modifications. More precisely,
      \begin{itemize}
      \item if the watermark $w$ is of the form $w=1 1^{\ell} 0 1^r b_n$, where $r-\ell=1$ and $b_n=0$, then $max\{{\tt minVM}(w)\}=4+\ell$ and the number of of true-watermarks $w'$ is $l+3$. By Theorem~\ref{thm:main}, Subcase 2.2, there is 1 true-watermark $w'$ if we apply {\tt Swap()}, other 1 if we apply {\tt Move in} and minimizing the value of $6+\ell+r-i$, and other $\ell+1$ true-watermark $w'$ if we apply {\tt Move-out} with $j \in [1,r]=[1,\ell+1]$.
      \item if the watermark $w$ is of the form $w=1 1^{\ell} 0 1^r b_n$, where $r-\ell=1$ and $b_n=1$, then $max\{{\tt minVM}(w)\}=4+\ell$ and the number of of true-watermarks $w'$ is $l$. By Theorem~\ref{thm:main}, Subcase 2.3, there is only 1 true-watermark $w'$ if we apply {\tt Swap()}.
      \item if the watermark $w$ is of the form $w=1 1^{\ell} 0 1^r b_n$, where $\ell - r = 1$ and $b_n=1$, then $max\{{\tt minVM}(w)\}=4+r$ and the number of of true-watermarks $w'$ is $l+3$. By Theorem~\ref{thm:main}, Subcase 2.3, there is $\ell$ true-watermark $w'$ if we apply {\tt Move-out} with $j \in [1,r+1]=[1,\ell]$.
      \end{itemize}
      We can see that the integer watermark $w$ with the minimum number of true-watermark $w'\in R_n$, where $n=2\kappa,\kappa \in \mathbb{Z}$, performing the maximum value of minimum valid edge-modifications is of the form $w=1 1^{\ell} 0 1^r b_n$, where $r-\ell=1$ and $b_n=1$. \ $\blacksquare$
\end{enumerate}

\begin{definition}
A watermark $w\in R_n$ is called {\it weak} if it has $minVM(w)=3$ in the range $R_n$.
\end{definition}

\noindent The weak watermarks $w$ are all watermarks numbers, that the internal block $B$ of their binary representation contains at least two 0s (see, Theorem~\ref{thm:main}).

\end{document}